\documentclass[twocolumn]{article}
\usepackage{subfiles}
\usepackage{graphicx}
\usepackage{amsmath,amssymb,amsfonts}
\usepackage{algorithmic}
\usepackage{textcomp}
\usepackage{xcolor}
\usepackage{soul}
\usepackage{url}
\usepackage{authblk}
\usepackage[labelfont=bf]{caption}
\usepackage[capitalise]{cleveref}
\usepackage[labelformat=simple]{subcaption}

\usepackage[numbers,sort&compress]{natbib}

\DeclareMathOperator{\sgn}{sgn}
\newcommand{\meanabs}{\langle |s_m| \rangle}

\begin{document}
\title{How to Incorporate Higher-order Interactions\\in Analog Ising Machines}

\author[1,2]{Robbe De Prins}
\author[2]{Guy Van der Sande}
\author[1]{Peter Bienstman}
\author[3]{Thomas Van Vaerenbergh}

\affil[1]{Photonics Research Group, Ghent University – imec, Technologiepark-Zwijnaarde 126, 9052 Gent, Belgium}
\affil[2]{Applied Physics Research Group, Vrije Universiteit Brussel, Pleinlaan 2, 1050 Brussels, Belgium}
\affil[3]{Large-Scale Integrated Photonics Lab, Hewlett Packard Labs, HPE Belgium, Diegem, Belgium}

\date{}
\maketitle

\begin{abstract}
Ising machines (IMs) are specialized devices designed to efficiently solve combinatorial optimization problems. Among such problems, Boolean Satisfiability (SAT) is particularly relevant in industrial applications. To solve SAT problems using IMs, it is crucial to incorporate higher-order interactions. However, in analog IMs, interactions of different orders scale unevenly with the continuous spin amplitudes, introducing imbalances that can significantly degrade performance. We present a numerical comparison of methods to mitigate these imbalances, evaluating time-to-solution and success rate on Uniform Random 3-SAT instances from the SATLIB benchmark set. Our results show that the most effective approach employs spin interactions that are proportional to the signs of spins, rather than their continuous amplitudes. This generalizes our previous work, which showed that such interactions best mitigate imbalances induced by external fields in quadratic analog IMs. In this work, its advantage becomes substantially more pronounced, as it naturally mitigates imbalances across all interaction orders. We further demonstrate that smooth approximations of this method make it compatible with analog hardware. Our findings underscore the central role of spin-sign-based interactions in enabling robust and scalable analog IM dynamics.
\end{abstract}

\section{Introduction}
\label{sec:intro}

A broad spectrum of computational issues in science and industry can be formulated as combinatorial optimization problems (COPs). Among these, Boolean Satisfiability (SAT) problems are especially relevant to real-world industrial applications such as scheduling \cite{horbach2012using,gomes1998randomization}, planning \cite{kautz1996pushing,russell2016artificial}, software and hardware verification \cite{burch1992symbolic,biere1999symbolic,velev2001effective}, automatic test pattern generation \cite{larrabee1993explorations,stephan2002combinational}, and neurosymbolic AI \cite{trinh2024solving,yang2024fine}. 

SAT’s practical importance is paralleled by its foundational complexity: it is widely believed that no algorithm can consistently solve all instances in polynomial time. In fact, SAT was the first problem proven to be NP-complete, as established by the Cook–Levin theorem \cite{cook1971proc,balbach2023cook}. 
Consequently, a variety of heuristic and specialized methods have been developed to address SAT problems as efficiently as possible. Among these, Ising machines (IMs) have emerged as a class of promising candidates. 
These specialized physical or digital systems solve COPs by encoding them into an Ising model: a spin system governed by the Hamiltonian
\begin{equation}
\mathcal{H}\left(\boldsymbol{\sigma}\right)=-\sum_i^N J^{(1)}_i \sigma_i - \sum_{i<j}^N J^{(2)}_{ij} \sigma_i \sigma_j,
\label{eq:hamiltonian quadratic Ising}
\end{equation}
where $\sigma_i = \pm 1$ $(1 \leq i \leq N)$ denote the spins, and $J^{(1)}_i$ and $J^{(2)}_{ij}$ are the strengths of the external fields and pairwise couplings, respectively. The goal of the IM is to evolve toward low-energy configurations, which correspond to optimal or near-optimal solutions of the encoded COP.

Numerous IM implementations have been introduced to date \cite{OverviewMcMahon}. For the purpose of this paper, we categorize them into two classes: IMs that are built using intrinsically binary spins $\sigma_i=\pm1$ \cite{Probabilistic_computing_with_pbits,Conti,cai2020power,Quantum_annealing_DWave}, and IMs employing continuous variables $s_i \in \mathbb{R}$ (e.g. intensities, voltages, etc. )\cite{100000SpinsCIM,leleu2019destabilization,inspiration_idea_thomas,Poor_mans_CIM,Ising_machine_based_on_networks_of_subharmonic_electrical_resonators,jiang2023efficient,PolaritonCondensates_2}. Whereas the first group includes the bistability of the spins as a \textit{hard constraint}, the latter group, which we will further refer to as analog IMs, implements this as a \textit{soft constraint} \cite{meseguer2006soft}. 

To solve SAT problems directly on IMs, it is essential to incorporate \textit{higher-order spin interactions} beyond the quadratic couplings of \cref{eq:hamiltonian quadratic Ising}.
In this work, we focus on 3-SAT problems, a canonical and prominent model for hard SAT instances, which naturally map to higher-order Ising Hamiltonians of the form:
\begin{equation}
\mathcal{H}\left(\boldsymbol{\sigma}\right)=-\sum_i^N J^{(1)}_i \sigma_i - \sum_{i<j}^N J^{(2)}_{ij} \sigma_i \sigma_j - \sum_{i<j<k}^N J^{(3)}_{ijk} \sigma_i \sigma_j \sigma_k,
\label{eq:hamiltonian higher order}
\end{equation}
where the coefficients $J^{(p)}_{i_1 \ldots i_p}$ represent real-valued interaction strengths of order $p$.
In the past, it has been shown that techniques that reduce higher-order terms to quadratic and linear interactions drastically impair the solver's performance \cite{dobrynin2024energy,bybee2023efficient,valiante2021computational,hizzani2024memristor,pedretti2025solving}, underscoring the importance of incorporating these interactions natively.

However, incorporating higher-order spin interactions in \textit{analog} IMs can be challenging. These difficulties arise from the dynamics governing the time evolution of the IM. For 3-SAT problems in particular, we will show in the next section that the time evolution of a spin amplitude $s_i$ depends on terms of the form
\begin{equation}
    J_i^{(1)} + \sum_{j} J_{ij}^{(2)} s_j + \sum_{j<k} J_{ijk}^{(3)} s_j s_k.
\end{equation}
Because interactions of different (polynomial) orders scale unevenly with the continuous spin amplitudes, their relative magnitudes can become unbalanced.
In particular, if $s_i \ll 1$, the linear terms dominate the quadratic ones, which in turn dominate the cubic terms.
This effect generalizes the imbalances between spin couplings and external fields observed in quadratic IMs \cite{original_meanAbsTrick_paper,original_auxTrick_paper,mastiyage2023mean}, which we have shown to degrade the IM's performance in prior work \cite{deprins2025ExternalFields}. Here, this effect is amplified by the simultaneous presence of multiple interaction orders.

In this paper, we present a numerical comparison of methods aimed at mitigating such imbalances. We draw inspiration from existing analog higher-order IMs such as PolySimCIM \cite{PolySimCIM}, Hopf oscillator networks \cite{bybee2023efficient}, and the higher-order simulated bifurcation algorithm \cite{HO_simBif}. 
Where possible, we also extend techniques that are common to address imbalances between external fields and quadratic spin interactions \cite{original_meanAbsTrick_paper,original_auxTrick_paper}.
To ensure a fair comparison, all methods are consistently implemented on an IM with identical, well-defined nonlinear dynamics that enforce spin bistability (details are provided in the next section) and with gradient-based spin dynamics (i.e.~without momentum).

We benchmark the methods on Uniform Random 3-SAT instances from the SATLIB library \cite{satlib}, evaluating them  in terms of time-to-solution (TTS) and success rate (SR). We find that spin-sign-based interactions—first introduced in discrete simulated bifurcation \cite{inspiration_idea_thomas}—consistently yield superior performance for incorporating higher-order terms. These findings align with our previous work on quadratic Ising models \cite{deprins2025ExternalFields}. 

While a previous study on the higher-order simulated bifurcation algorithm \cite{HO_simBif} reported reduced performance when applying the spin sign method, we hypothesize that this outcome stems from their exclusive focus on a purely third-order model (without quadratic and linear terms), where no imbalances arise between interaction orders. In a more realistic setting for SAT problems, where interactions of multiple orders are included, we show that spin-sign-based interactions demonstrate a clear advantage, even more pronounced than in the setting with only quadratic couplings and external fields. 
Their ability to mitigate imbalances across all interaction orders highlights their critical role in designing effective analog IM dynamics.
We reinforce this conclusion by demonstrating that these interactions can be implemented on analog hardware through smooth approximations.

\section{Modeling analog Ising machines with higher-order spin interactions}
\label{sec:modeling IMs}

Before introducing various techniques to incorporate higher-order spin interactions, we describe the analog IM that is used to benchmark them. The time evolution of spin amplitude $s_i \in \mathbb{R}$ is modeled as:
\begin{equation}
    \frac{ds_i}{dt} = -s_i + \tanh\left(\alpha s_i + \beta I_i\right),
    \label{eq: sigmoid nonlin}
\end{equation}
where $\alpha$ is the linear gain and $\beta$ interaction strength. $\beta$ follows a commonly used linear annealing scheme \cite{paper_Ganguli,PaperJacob_UsingContinuationMethods} (see Methods for details). 
$I_i$ represents the local field acting on $s_i$, whose exact form depends on how interactions of different orders are incorporated (this is discussed in the following section).

We use the hyperbolic tangent nonlinearity due to its strong empirical performance in previous work \cite{Order_of_magnitude}, which can be attributed to the suppression of amplitude inhomogeneity, a common source of performance degradation in IMs. Moreover, it reflects the saturation behavior that is commonly observed in experimental IMs \cite{toonIISM2025}. 

The IM is simulated by numerically integrating \cref{eq: sigmoid nonlin} via the Euler–Maruyama method, which includes stochastic noise (see Methods for details).
At any time during the simulation, the system’s energy, as defined by the higher-order Hamiltonian of \cref{eq:hamiltonian higher order}, can be evaluated by mapping the continuous spin amplitudes to binary spin values using a sign function: $\sigma_i = \text{sgn}(s_i)$.

\subsection{Methods to incorporate higher-order interactions}
\label{sec:local_field_methods}
We compare five approaches for incorporating higher-order interactions into analog IMs, each specifying a distinct form for the local field $I_i$ in \cref{eq: sigmoid nonlin}. These methods are applied to 3-SAT problems of the form of \cref{eq:hamiltonian higher order}, but can be naturally extended to problems with interactions beyond third order.

\subsubsection*{Baseline method}

As a baseline, we consider the following local field:
\begin{equation}
    I_i = J_i^{(1)} + \sum_{j} J_{ij}^{(2)} s_j + \sum_{j<k} J_{ijk}^{(3)} s_j s_k.
\label{eq:baseline}
\end{equation}
This expression, also used in PolySimCIM \cite{PolySimCIM} and in Coherent SAT Solvers \cite{reifenstein2023coherent}, is a direct extension of the typical local field used, for example, in Coherent Ising machines \cite{100000SpinsCIM}. 
It can be deduced from \cref{eq:hamiltonian higher order} by relaxing the binary spins $\sigma_i$ to continuous variables $s_i \in \mathbb{R}$ and taking the negative of the partial derivative with respect to $s_i$.

When programming the IM, the coefficients $J^{(p)}_{i_1 \ldots i_p}$ are carefully chosen to ensure that \cref{eq:hamiltonian higher order} encodes the intended COP (i.e.~its ground state corresponds to the solution). 
However, this desired weighting among terms disappears when the spin amplitudes in \cref{eq:baseline} differ significantly from unity.
In particular, if $s_i \ll 1$, the linear terms dominate the quadratic ones, which in turn dominate the cubic terms. Since such imbalances effectively discard information that is crucial to solving the COP, they can degrade the IM's performance. Note that the opposite case, where $s_i \gg 1$, is excluded, since we employ the tanh-nonlinearity of \cref{eq: sigmoid nonlin}, which restricts the spin amplitudes within [-1,1]. 

\subsubsection*{Rescaling method 1}
As a first way to address the imbalances between interaction orders, we propose the following expression:
\begin{equation}
I_i=J_i^{(1)}+\sum_j J_{i j}^{(2)} \frac{s_j}{\meanabs}+\sum_{j<k} J_{i j k}^{(3)} \frac{s_j s_k}{\meanabs^2},
\label{eq:rescaling_method1}
\end{equation}
where $\meanabs$ is the average absolute value of all spin amplitudes in the IM.
This approach generalizes a technique originally proposed for purely quadratic IMs \cite{original_meanAbsTrick_paper}. To the best of our knowledge, this extension has not been reported previously. By rescaling the spin amplitudes with $\meanabs^{(-1)}$ and $\meanabs^{(-2)}$, the scaling of the quadratic and cubic terms is adjusted to align with that of the linear terms $J^{(1)}_i$, preserving the intended balance among contributions.

\subsubsection*{Rescaling method 2}
Alternatively, one can align the scaling of the linear and cubic terms with that of the quadratic couplings ($\propto J^{(2)}_{ij}$):
\begin{equation}
    I_i = J^{(1)} \meanabs + \sum_j J_{ij}^{(2)} s_j + \sum_{j<k} J_{ijk}^{(3)} \frac{s_j s_k}{\meanabs}.
\label{eq:rescaling_method2}
\end{equation}
This expression is obtained from \cref{eq:rescaling_method1} by multiplying the entire local field by a global factor $\meanabs$.

\subsubsection*{Rescaling method 3}
By multiplying \cref{eq:rescaling_method2} once more by $\meanabs$, we obtain:
\begin{equation}
    I_i = J_i^{(1)}\meanabs^2 + \sum_j J_{ij}^{(2)} s_j \meanabs + \sum_{j < k} J_{ijk}^{(3)} s_j s_k,
\label{eq:rescaling_method3}
\end{equation}
which aligns the scaling of the linear and quadratic terms with the cubic terms ($\propto J^{(3)}_{ijk}$).

\subsubsection*{Spin sign method}
Finally, we consider the following local field expression:
\begin{equation}
    I_i = J_i^{(1)} + \sum_{j} J_{ij}^{(2)} \sgn(s_j) + \sum_{j<k} J_{ijk}^{(3)} \sgn(s_j) \sgn(s_k).
\label{eq:spinsign}
\end{equation}
It was originally proposed in the context of discrete simulated bifurcation \cite{inspiration_idea_thomas,HO_simBif}. A similar approach for complex-valued spins represented by Hopf oscillators is proposed in Ref.~\cite{bybee2023efficient}.
In our previous work \cite{deprins2025ExternalFields}, we showed that this approach is effective at mitigating imbalances caused by external fields in quadratic IMs.

\section{Results}
\label{sec:results}

\subsection{Performance comparison}
We benchmark the methods for incorporating higher-order interactions from \cref{sec:local_field_methods} on Uniform Random-3-SAT problems from the SATLIB library (see Methods for details).
Specifically, we consider the first 10 instances of each of the following SATLIB subsets: uf20-91, uf50-218, uf100-430, uf150-645, uf200-860, and uf250-1065. 
All of these instances are satisfiable, meaning the ground-state energy—used as the target for the TTS and SR metrics—corresponds to the situation where all boolean clauses are satisfied. As detailed in Methods, these performance metrics are computed by optimizing over the relevant hyperparameters of the IM.

\begin{figure*}
    \includegraphics[width=1.0\linewidth]{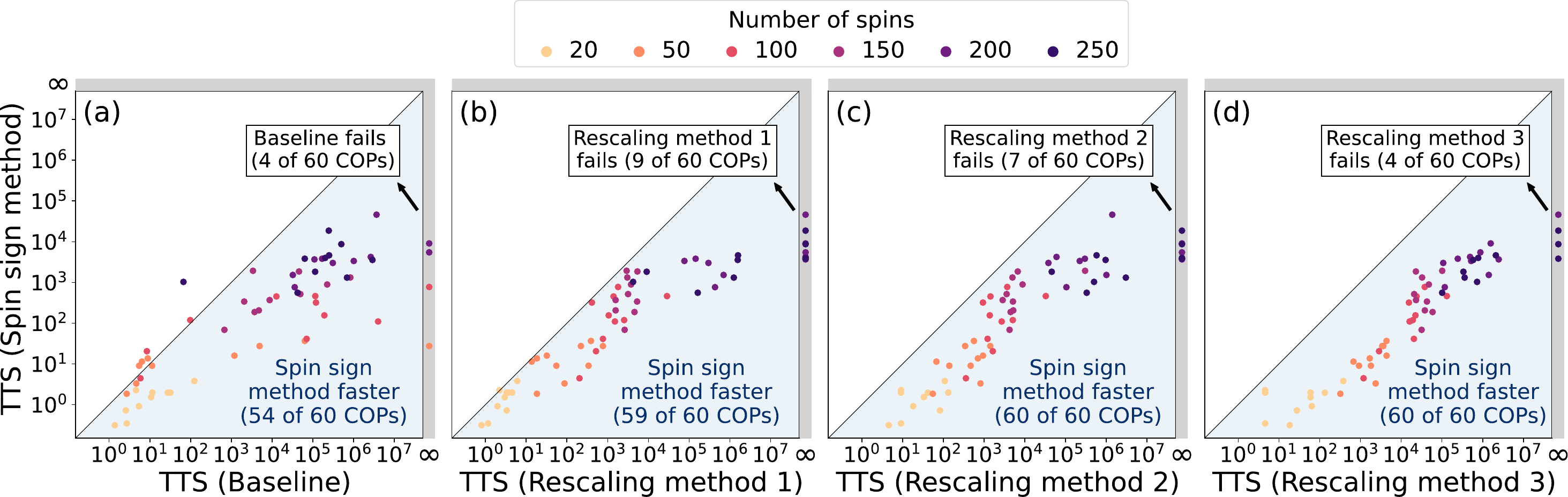}
    \caption{Comparison of TTS between the methods to incorporate higher-order interactions of \cref{sec:local_field_methods} for Uniform Random-3-SAT problems. 
    Dots in the grey area on the right denote COPs that could be solved by the spin sign method within the allocated compute time of $t_\text{max}=10^4$, but not by the method on the x-axis ($\text{TTS}=\infty$, $\text{SR}=0$).
    The spin sign method  generally requires less time to solve problems, and solves more problems within the allocated time.
    Details about the simulation procedure are provided in Methods. 
    }
    \label{fig: TTS comparison}
\end{figure*}

\cref{fig: TTS comparison} presents a pairwise comparison of the methods from \cref{sec:local_field_methods} using the TTS metric. Each point represents a problem instance, with the $y$-value denoting the TTS of the spin sign method from \cref{eq:spinsign}, and the $x$-value denoting the TTS of the reference method.

\cref{fig: TTS comparison}(a) compares the spin sign method with the baseline method of \cref{eq:baseline}. 
54 of the 60 COPs lie below the diagonal, indicating that they are solved faster using the the spin sign method than using the baseline method. 
Notably, 4 of these 54 points fall in the grey region on the right, where $\text{TTS}=\infty$, meaning the baseline method could not solve them for any of the attempted hyperparameter values (see Methods for details) within the time limit $t_\text{max}=10^4$. In contrast, the spin sign method successfully solved all four, yielding finite TTS values. 

\cref{fig: TTS comparison}(b) contrasts the spin sign method with rescaling method 1 of \cref{eq:rescaling_method1}. In this comparison, the spin sign method outperforms on 59 of the 60 problems, with rescaling method 1 failing to solve 9 of them. The single instance where the spin sign method is slower lies nearly on the diagonal, indicating comparable performance.

\cref{fig: TTS comparison}(c) and (d) compare the spin sign method with rescaling methods 2 and 3, respectively. The spin sign method outperforms both across all instances. Rescaling method 2 fails to solve 7 of them, while rescaling method 3 fails on 4. Looking at \cref{fig: TTS comparison}(b-d) combined, we see a consistent pattern among the rescaling strategies. As seen from \cref{eq:rescaling_method1,eq:rescaling_method2,eq:rescaling_method3}, these models differ only by a global scaling factor $\meanabs^q$ with $q \in \mathbb{Z}$. Increasing $q$ (i.e.~moving from method 1 to 2 to 3) tends to increase the average TTS for smaller instances (lighter dots), but improves performance on larger ones (darker dots), as reflected by a decrease in the number of unsolved problems ($\text{TTS} = \infty$). A possible explanation is that multiplying the local fields $I_i$ by $0 \leq \meanabs \leq 1$, and thus increasing $q$, slows down the IM dynamics, which may help avoid premature convergence in larger instances (which are generally harder) but becomes unnecessary overhead for smaller (generally easier) ones. 

Combining the information from all panels from \cref{fig: TTS comparison}, the spin sign method shows clear advantages over the other four methods: it manages to solve more problems, and it generally yields lower TTS values. Notably, this advantage becomes more pronounced with increasing problem size. This trend is further confirmed in \cref{fig:TTS_comparison_boxplots} of the Supplementary Material, which shows the TTS distributions for all methods as a function of problem size.

Detailed per-instance results underlying \cref{fig: TTS comparison} are provided in \cref{fig: TTS comparison individual} in the Supplementary Material. Additionally, while \cref{fig: TTS comparison} uses a timestep of $dt=0.01$ (see Methods), \cref{fig:smaller_timestep} presents a similar TTS analysis with a smaller timestep of $dt=0.001$. The close agreement between the two confirms that $dt=0.01$ is sufficiently small to accurately integrate the IM’s governing equations (\cref{eq: sigmoid nonlin}) for the problems that we studied.

\begin{figure*}
    \includegraphics[width=1.0\linewidth]{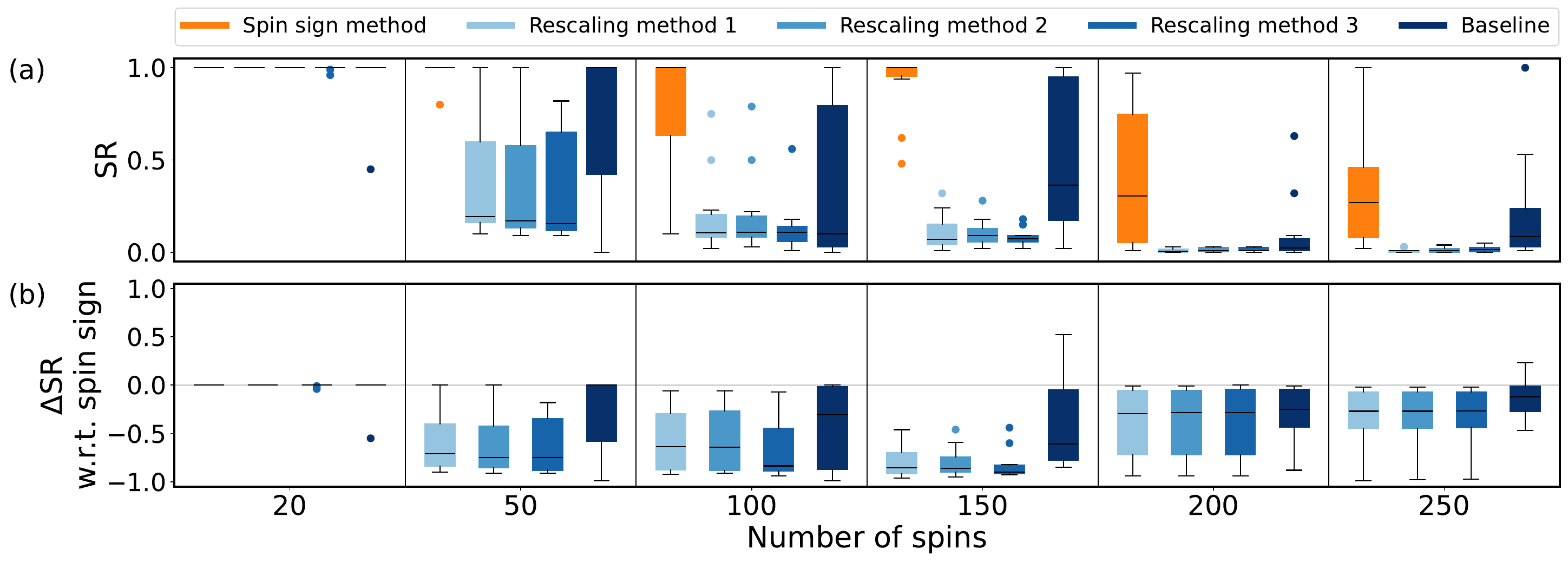}
    \caption{Comparison of success rates across the different methods to incorporate higher-order interactions of \cref{sec:local_field_methods} on Uniform Random-3-SAT problems. Panel (a) shows absolute SRs, while panel (b) displays SRs relative to the spin sign method. Each boxplot represents a distribution over 10 instances of the same size for a given method, where dots indicate outliers. A detailed breakdown for individual problem instances is provided in \cref{fig: SR comparison individual} of the Supplementary Material. 
    }
    \label{fig: SR comparison boxplot}
\end{figure*}

We now confirm the advantage of the spin sign method over its alternatives using a second performance metric: the success rate. Instead of evaluating $\min_\Theta \text{TTS}$, we now consider $\max_\Theta \text{SR}$, where $\Theta$ denotes the set of hyperparameters (see Methods for details). \cref{fig: SR comparison boxplot}(a) compares the SR of all methods across problem sizes. Each boxplot represents the SR distribution over 10 problem instances of the same size. We observe that all methods succeed at solving the problems of size 20 with near-perfect SR. However, as the problem size increases, the SR generally decreases. This decrease is slower for the spin sign method than for its alternatives.
\cref{fig: SR comparison boxplot}(b) visualizes the SR values relative to the spin sign method. We see that for almost all problems, the spin sign method yields equal or higher SR values than all other methods.
Instance-specific data is provided in \cref{fig: SR comparison individual} of the Supplementary Material. 

\subsection{Approximate spin sign method: simulating hardware-induced effects}

When using the spin sign method of \cref{eq:spinsign}, the continuous spin amplitudes are mapped to their signs using the exact (discontinuous) sign function when computing local fields. While this works in simulation, physical implementations in analog hardware inevitably require a smooth approximation with finite steepness. To explore how this hardware-induced constraint affects performance, we introduce and evaluate a continuous approximation of the sign function:
\begin{equation}
    \sgn(s_i) \approx \tanh(\kappa s_i),
\end{equation}
where the steepness parameter $\kappa \in \mathbb{R}$ controls the sharpness of the transition. For large $\kappa$, the approximation closely matches the exact sign function; smaller values yield more gradual transitions.

\begin{figure}
    \includegraphics[width=1.0\linewidth]{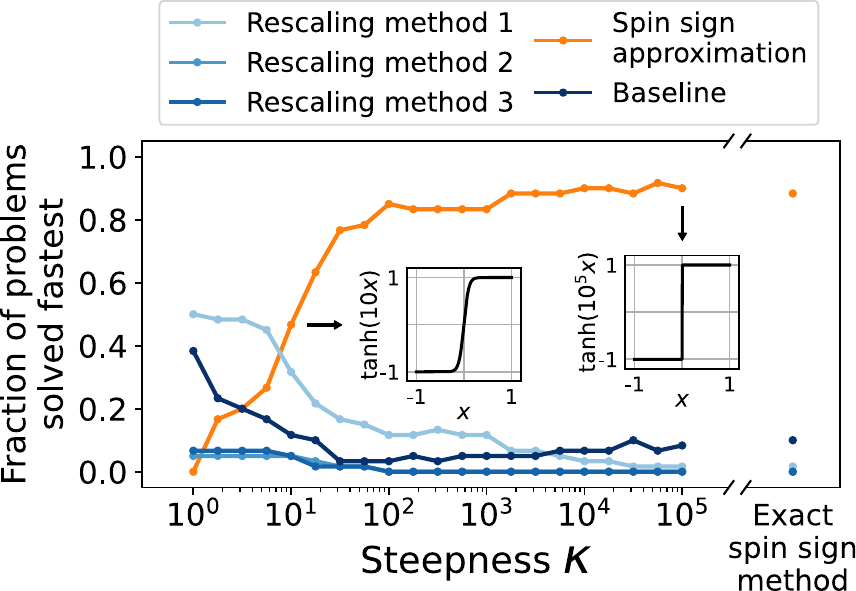}
    \caption{
    Fraction of benchmark problems for which each method is fastest, as a function of $\kappa$ in the smooth $\tanh(\kappa s_i)$ approximation of the spin sign method. As the performance of the approximate method improves with increasing $\kappa$, the relative ranking across all methods shifts, eventually reaching the limit of the exact spin sign method.
    }
    \label{fig: rank plot spinsign approx}
\end{figure}

For varying values of $\kappa$, \cref{fig: rank plot spinsign approx} shows the fraction of the benchmark problems that are solved fastest using each of the local field methods from \cref{sec:local_field_methods}. The curves sum to 1 at every value of $\kappa$. While only the approximate spin sign method depends on $\kappa$, changes in its performance affect the relative ranking of all methods. At $\kappa = 1$, we observe that the approximation is too smooth to be effective, and the method fails to outperform any alternative. As $\kappa$ increases, the approximate spin sign method quickly improves. At $\kappa = 10$, it becomes the fastest method on the largest fraction of problems. As $\kappa$ increases further, the performance converges toward that of the exact spin sign method.

\section{Conclusion}
We have benchmarked five methods for incorporating higher-order spin interactions into analog Ising machines. Our results show that the choice of local field formulation critically impacts performance, especially as problem size increases. While strategies that rescale the spin amplitudes with the mean absolute value of all spins partially mitigate amplitude imbalance and improve performance over the naive baseline, they are consistently outperformed by the spin sign method. This last method achieves higher success rates and lower time-to-solution across a wide range of problem sizes. Its performance also scales more favorably than the other methods, making it increasingly effective for larger problems.

These findings underscore the importance of maintaining balance among interaction terms for effective problem solving. Enforcing binary behavior in spin contributions to the local field helps preserve the problem’s discrete structure, leading to clear performance gains even in continuous-state dynamical systems. 
This serves as a key factor in advancing analog Ising machine architectures.

Beyond performance, the spin sign method offers notable practical advantages. As discussed in more detail in our previous work \cite{deprins2025ExternalFields}, this method is well-suited for hardware implementation. In many experimental IMs, spin amplitudes are already measured before computing local fields \cite{2000nodes,100000SpinsCIM,Poor_mans_CIM,Ising_machine_based_on_networks_of_subharmonic_electrical_resonators,jiang2023efficient,PolaritonCondensates,PolaritonCondensates_2}, such that a simple one-bit comparator suffices to extract the spin sign. Whereas obtaining the sign of the spins is a local operation, calculating the average absolute spin value is a global operation over all spins, which is less trivial to implement.
Additionally, in this work, we showed that smooth approximations of the spin sign method, which arise naturally in analog hardware, can match the performance of the exact method, provided the approximation is sufficiently accurate.

This work focused on the local field formulation for higher-order interactions, keeping the dynamics of the analog IM fixed (cf.~\cref{eq: sigmoid nonlin}). Future work could explore how modifying the dynamics—such as adding momentum terms to the gradient-based spin updates \cite{inspiration_idea_thomas,HO_simBif} or incorporating chaotic amplitude control \cite{Destabilization_of_local_minima,Scaling_advantage_of_chaotic_amplitude_control,reifenstein2023coherent} — might  further enhance performance. It would also be interesting to extend this comparison with higher-order IMs that employ phase variables, similar to the Kuramoto model \cite{bashar2023designing}.

Finally, comparing our approach against state-of-the-art SAT solvers is an important next step. 
While the direct TTS comparisons presented here effectively evaluate the local field methods from \cref{sec:local_field_methods} within the analog IM framework, broader comparisons across fundamentally different algorithms and hardware require implementation-agnostic metrics. To this end, we can estimate the empirical time complexity of our solver \cite{mu2015empirical,mu2015analysing}, which involves extensive benchmarking over a wider range of problem sizes and instance sets. Pursuing these directions will help assess the full potential of analog IMs for solving higher-order combinatorial problems.

\section{Methods}
\label{sec:methods}

\subsection{Embedding SAT problems in a higher-order Ising model}
\label{sec:methods:embedding SAT}

The goal of a SAT problem is to find an assignment to a set of $N$ Boolean variables such that a given logical formula evaluates to \texttt{True}.
Every SAT problem can be expressed in conjunctive normal form (CNF) as a conjunction (logical AND, $\land$) of clauses, each of which is a disjunction (logical OR, $\lor$) of exactly $L$ literals. A literal $\texttt{l}$ is either a variable $\texttt{x}_i$ or its negation $\lnot \texttt{x}_i$. In the case of 3-SAT, we have $L=3$.

A 3-SAT problem with $C$ clauses can be expressed as:
\begin{equation}
    \bigwedge_{i=1}^C \left( \texttt{l}_{i1} \lor \texttt{l}_{i2} \lor \texttt{l}_{i3} \right),
\end{equation}
where $\texttt{l}_{ij}$ denotes the $j$-th literal in the $i$-th clause.

To embed a 3-SAT problem into the higher-order Ising Hamiltonian defined in \cref{eq:hamiltonian higher order}, we first convert the logical formula into a real-valued PUBO energy function of binary variables $x_k\in\{0,1\}$:
\begin{equation}
    H(\mathbf{x}) = \sum_{i=1}^C (1 - l_{i1}) (1 - l_{i2}) (1 - l_{i3}),
    \label{eq:sat_mapping_pubo}
\end{equation}
where
\begin{equation}
l_{ij} = 
\begin{cases}
x_k & \text{if the literal } \texttt{l}_{ij} \text{ is } \texttt{x}_k, \\
1 - x_k & \text{if the literal } \texttt{l}_{ij} \text{ is } \neg \texttt{x}_k,
\end{cases}
\end{equation}
and $k\in\{1,\dots,N\}$. Note that every unsatisfied clause adds an energy penalty of $+1$ to \cref{eq:sat_mapping_pubo}. 

We then express the energy function in terms of Ising spins $\sigma_i=\pm1$ using the following transformation:
\begin{equation}
    x_k = \frac{\sigma_k+1}{2}.
    \label{eq: transfo bits to spin}
\end{equation}

\subsection{Simulating the analog Ising machine}
The temporal evolution of the Ising machine is modeled using the Euler–Maruyama integration method:
\begin{equation}
\mathbf{s}_{t+1} = \mathbf{s}_t + \Delta t \left(-\mathbf{s}_t + \tanh\left(\alpha \mathbf{s}_t + \beta_t \mathbf{I}_t\right)\right) + \gamma \boldsymbol{\xi}_t,
\label{eq:euler_update}
\end{equation}
where $\mathbf{s}_t$ denotes the vector of spin amplitudes at time $t$, $\alpha$ is the linear gain, and $\Delta t = 0.01$ is the time step (which is validated as detailed in the Supplementary Material). The local fields acting on the spins at time $t$ are encoded in the vector $\mathbf{I}_t$, according to one of the methods of \cref{sec:local_field_methods}. $\gamma$ denotes the noise strength, while each noise component $\xi_{t,i}$ is sampled independently from a normal distribution with zero mean and standard deviation $\sqrt{\Delta t}$. The interaction strength $\beta_t$ is gradually increased over time using:
\begin{equation}
    \label{eq:linear_annealing}
    \beta_{t+1} = \beta_t + v_\beta\Delta t,
\end{equation}
starting from $\beta_0 = 0$, with $v_\beta$ denoting the annealing rate.

The initial values of the spin amplitudes are sampled from the same Gaussian distribution as the noise process, i.e.~$\gamma \boldsymbol{\xi}_0$.
The iterative updates of \cref{eq:euler_update,eq:linear_annealing} continue until the IM either achieves the target energy or completes $10^4/\Delta t$ steps.

For each problem instance, we perform a grid search over the hyperparameters, as detailed below. Each of the reported TTS values (e.g.~in \cref{fig: TTS comparison}) corresponds to the optimal hyperparameter configuration identified through this process. For each hyperparameter setting, the IM evolution is repeated 100 times to estimate the TTS.
\begin{itemize}
\item Linear gain: $\alpha \in \{-10, -9, \dotsc, 1\}$
\item Annealing speed: \\$v_\beta \in \{10^{-4}, 10^{-3}, 10^{-2}, 10^{-1}, 1\}$
\item Noise strength: $\gamma \in \{10^{-4}, 10^{-3}, 10^{-2}, 10^{-1}\}$
\end{itemize}

\section{Data availability}
The authors declare that all relevant data are included in the manuscript. Additional data are available from the corresponding author upon reasonable request.

\section{Author contributions}
R.D.P. performed the simulations and wrote the manuscript. G.V.d.S., P.B., and T.V.V. supervised the project. All authors discussed the results and reviewed the manuscript.

\section{Additional information}
{\bf Competing interests:} The authors declare no competing interests.\\
{\bf Acknowledgements:} 
We would like to thank Adrien Renaudineau for his work on implementing the PUBO embeddings of SAT problems. 
This research was funded by the Horizon Europe framework (Prometheus project 101070195) and by the Research Foundation Flanders (FWO) under grants G028618N, G029519N, G0A6L25N, and G006020N. Additional funding was provided by the EOS project `Photonic Ising Machines'. This project (EOS number 40007536) has received funding from the FWO and F.R.S.-FNRS under the Excellence of Science (EOS) programme. The work was also partly supported by the Defense Advanced Research Projects Agency (DARPA) under Air Force Research Laboratory (AFRL) contract no FA8650-23-3-7313.

\bibliographystyle{unsrtnat}
\bibliography{references}
\clearpage

\renewcommand{\theequation}{S.\arabic{equation}}
\renewcommand{\thetable}{S\arabic{table}}
\renewcommand{\thefigure}{S\arabic{figure}}
\renewcommand{\thesection}{S.\arabic{section}}
\setcounter{equation}{0}
\setcounter{figure}{0}
\setcounter{section}{0}

\twocolumn[
\vspace*{1cm}
\begin{center}
  {\LARGE \bfseries Supplementary Material}
\end{center}
\vspace{0.5cm}
]

\section{Success rates and time-to-solution for individual problem instances}

\begin{figure*}[htb]
    \centering
    \includegraphics[width=1.0\linewidth]{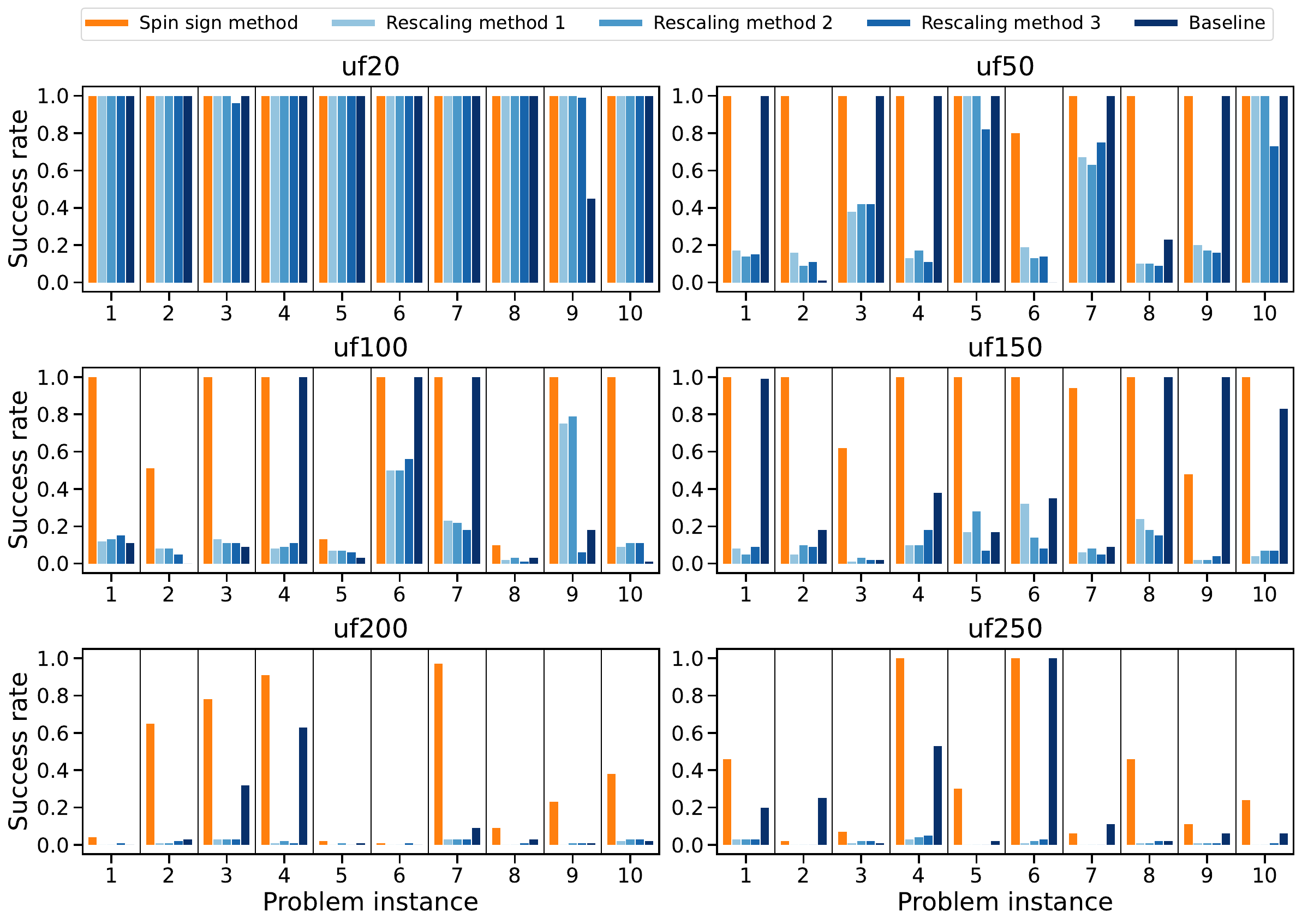}
    \caption{Instance-specific success rates for the Uniform Random-3-SAT instances, underlying the boxplots in \cref{fig: SR comparison boxplot}. We compare the various methods to incorporate higher-order interactions, as described in \cref{sec:local_field_methods}. Problem names follow the format \texttt{uf\{N\}-\{i\}}, where \texttt{N} is the number of spin variables (indicated in the subplot title) and \texttt{i} is the instance identifier (shown along the x-axis).
    }
    \label{fig: SR comparison individual}
\end{figure*}

\begin{figure*}[htb]
    \centering
    \includegraphics[width=1.0\linewidth]{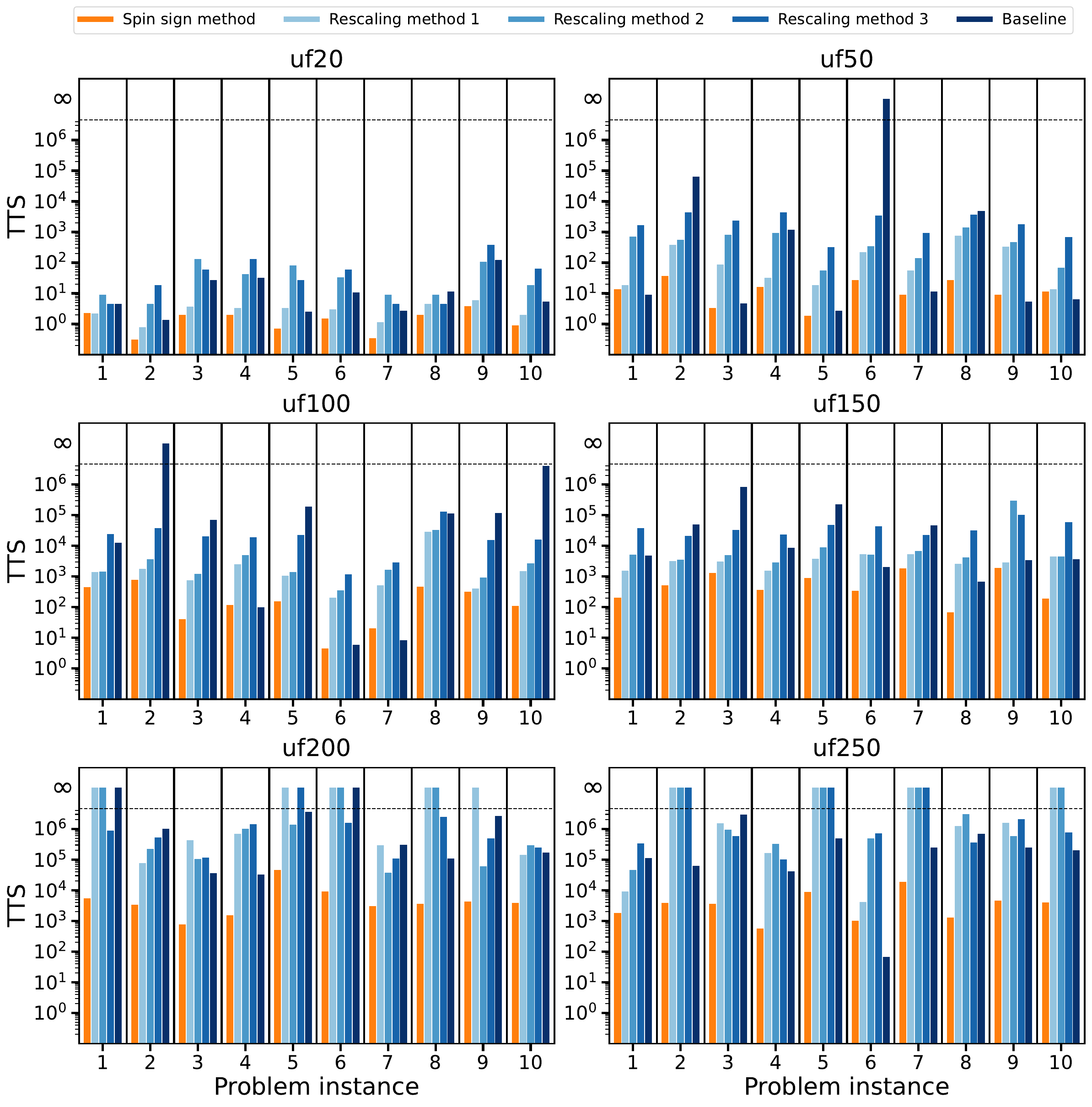}
    \caption{Instance-specific time-to-solution values for the Uniform Random-3-SAT instances, underlying \cref{fig: TTS comparison}. We compare the various methods to incorporate higher-order interactions, as described in \cref{sec:local_field_methods}. The dashed line denotes the resolution limit of the TTS metric. All bars surpassing this limit denote $\text{TTS}=\infty$, meaning they were not solved by the given method for any of the hyperparameter values (cf.~\cref{sec:methods}).  Problem names follow the format \texttt{uf\{N\}-\{i\}}, where \texttt{N} is the number of spin variables (indicated in the subplot title) and \texttt{i} is the instance identifier (shown along the x-axis).}
    \label{fig: TTS comparison individual}
\end{figure*}

\cref{fig: SR comparison individual} shows the success rate achieved on each individual Uniform Random-3-SAT instance, for each of the methods of \cref{sec:local_field_methods}. This breakdown highlights the variability across instances that is not visible in the aggregated boxplots of \cref{fig: SR comparison boxplot}.
Similarly, \cref{fig: TTS comparison individual} shows the TTS values for individual problems and local field methods, underlying \cref{fig: TTS comparison}.

\section{Distribution of time-to-solution across problem sizes}
\begin{figure*}
    \includegraphics[width=1.0\linewidth]{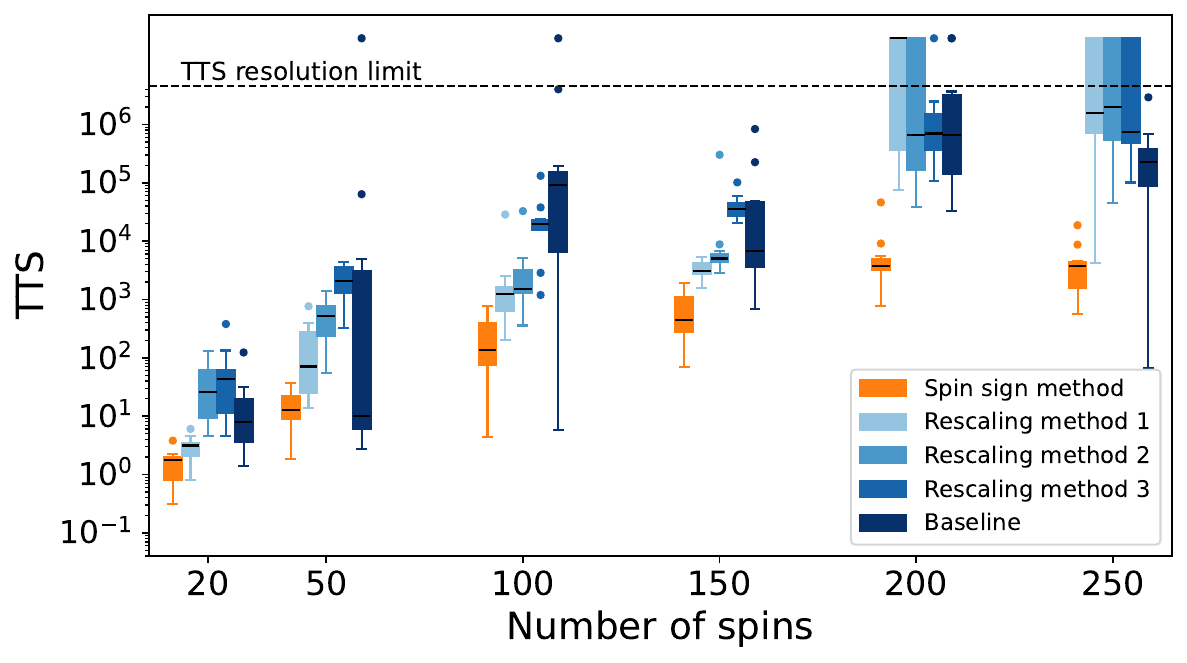}
    \caption{TTS distributions across problem sizes for all methods from \cref{sec:local_field_methods}. Each boxplot represents TTS values over 10 instances of the same size. Outliers are shown as dots. The advantage of the spin sign method over the alternative methods grows with increasing problem size.}
    \label{fig:TTS_comparison_boxplots}
\end{figure*}

In \cref{fig: TTS comparison} of the main text, we observed that data points increasingly diverge from the diagonal as problem size grows (indicated by darker dots). To make this trend more explicit, we show the TTS distribution for all methods from \cref{sec:local_field_methods}, grouped by problem size. As seen in \cref{fig:TTS_comparison_boxplots}, the spin sign method maintains lower TTS values than the alternative methods, with the performance gap widening as the number of spins increases.

\section{Validation of time step size}
\begin{figure*}[htb]
    \centering
    \includegraphics[width=1.0\linewidth]{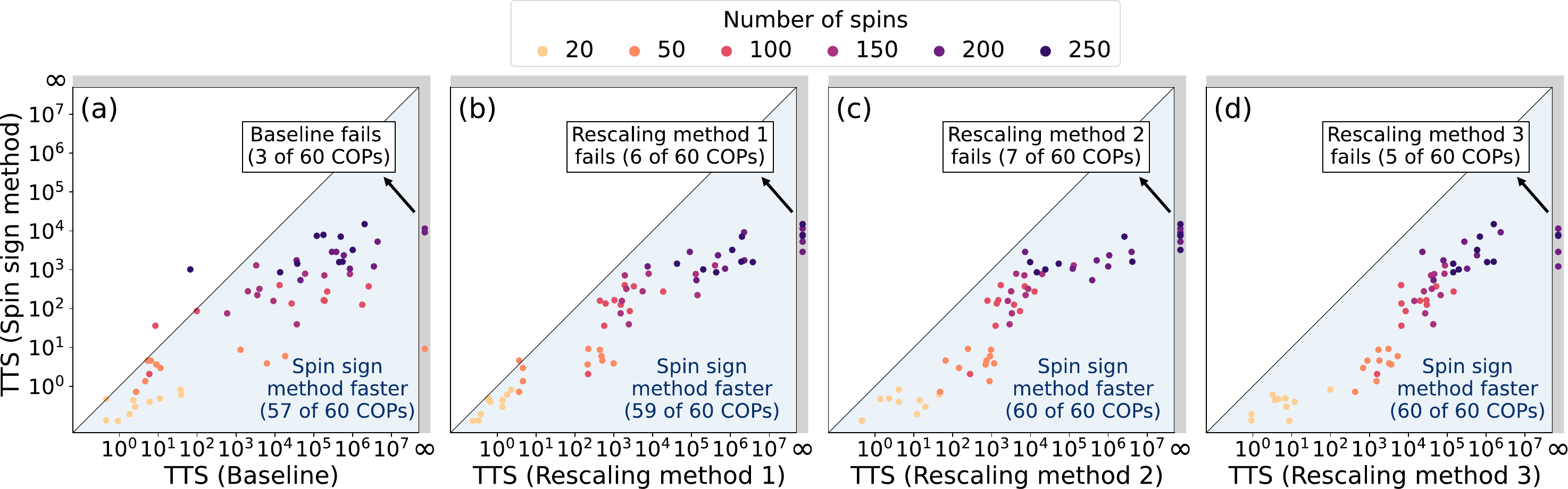}
    \caption{Comparison of TTS across methods for incorporating higher-order interactions (see \cref{sec:local_field_methods}) on Uniform Random-3-SAT instances. This figure mirrors \cref{fig: TTS comparison} from the main text but uses a smaller timestep of $dt=0.001$ instead of $dt=0.01$. The similarity with \cref{fig: TTS comparison} indicates that $dt=0.01$ is sufficient for accurately integrating the IM's governing equations (\cref{eq: sigmoid nonlin}).
    }
    \label{fig:smaller_timestep}
\end{figure*}

As described in \cref{sec:methods}, all results in the main text were obtained using a time step of $dt=0.01$. To ensure this choice is sufficiently small for accurately integrating the governing equations of the IM (\cref{eq: sigmoid nonlin}), we repeat the TTS analysis with a smaller time step of $dt=0.001$. The results, shown in \cref{fig:smaller_timestep}, closely match those in \cref{fig: TTS comparison}, confirming that $dt=0.01$ yields reliable integration.

\end{document}